%% file: main.tex
\documentclass[prd,aps,preprint,amsmath,amssymb,superscriptaddress]{revtex4-2}
\def\rem#1{ {\bf\textcolor{red}{($\spadesuit$ #1 $\spadesuit$)}}}

\usepackage{comment}

\input{cassX}%插入我常用的包和定义，给了一些标准的微分算子定义，

\begin{document}

%\preprint{}

\title{Testing the Transverse Scalar Mode of Gravitational Quantum Field Theory with Taiji and LISA}


\author{Cong Xu\,\orcidlink{0009-0009-7151-3886}}
\email{xucong22@mails.ucas.ac.cn}
%\thanks{}
\affiliation{School of Fundamental Physics and Mathematical Sciences, Hangzhou Institute for Advanced Study, UCAS, Hangzhou 310024, China}
\affiliation{Institute of Theoretical Physics, Chinese Academy of Sciences, Beijing 100190, China}
\affiliation{University of Chinese Academy of Sciences, Beijing 100049, China}

\author{Yong Tang\,\orcidlink{0000-0003-1100-2741}}
\email{tangy@ucas.ac.cn}
%\affiliation{School of Astronomy and Space Sciences, University of Chinese Academy of Sciences (UCAS), Beijing 100049, China}
\affiliation{School of Fundamental Physics and Mathematical Sciences, Hangzhou Institute for Advanced Study, UCAS, Hangzhou 310024, China}
\affiliation{University of Chinese Academy of Sciences, Beijing 100049, China}
\affiliation{International Centre for Theoretical Physics Asia-Pacific, UCAS, Beijing 100190, China}

\author{Yue-Liang Wu}
\email{ylwu@ucas.ac.cn}
\affiliation{School of Fundamental Physics and Mathematical Sciences, Hangzhou Institute for Advanced Study, UCAS, Hangzhou 310024, China}
\affiliation{Institute of Theoretical Physics, Chinese Academy of Sciences, Beijing 100190, China}
\affiliation{University of Chinese Academy of Sciences, Beijing 100049, China}
\affiliation{International Centre for Theoretical Physics Asia-Pacific, UCAS, Beijing 100190, China}
\affiliation{Taiji Laboratory for GW Universe (Beijing/Hangzhou), University of Chinese Academy of Sciences, Beijing 100049, China}


\date{\today}

\begin{abstract}
Space-based gravitational-wave (GW) detectors, including LISA and Taiji, offer unprecedented access to regimes where alternative theories of gravity may deviate from General Relativity (GR). Gravitational Quantum Field Theory (GQFT) provides a novel framework in which the Poincar\'e-type inhomogeneous spin symmetry of Weyl-type fermions in the Standard Model is elevated to a gauge symmetry. Within this construction, the fundamental gravitational field is identified with a gravigauge field which behaves as a Goldstone-type bi-covariant vector field. Unlike GR, GQFT predicts additional polarization states: one transverse scalar (breathing) mode and two vector modes. In this work, we focus on the transverse, isotropic scalar mode and investigate its detectability with Taiji. To isolate this mode, we employ the null-response channel (NRC), a specific interferometric combination designed to suppress contributions from other polarizations. We implement an analytical, dynamic orbital model to realistically simulate a triangular constellation. We compute the response functions and sensitivity curves for various interferometric channels, compare them with the standard Michelson channel, and demonstrate the effectiveness of the NRC approach. Our results show that the NRC provides a reliable, waveform-independent criterion for testing non-GR polarizations, and we anticipate that it will serve as a valuable tool for probing gravitational theories in future space-based GW missions.
\newpage
\end{abstract}

%\keywords{}

\maketitle


\section{Introduction}
A century after the formulation of general relativity (GR), gravitational waves (GWs) were first observed from a binary black hole merger through ground-based detection by LIGO/Virgo \cite{Castelvecchi2016,Abbott2016}, ushering in a new era for gravitational physics. 
GWs can penetrate regions inaccessible to electromagnetic waves, aiding our understanding of black hole mergers in the universe or the early universe before recombination, and more. 
Compared to ground-based observations constrained by Earth's environment and limited arm lengths, space-borne observatories like LISA \cite{Danzmann1996} and Taiji \cite{Hu2017} benefit from less interference and longer arms, extending their sensitivity to the millihertz frequency band. 

Both missions employ a constellation of three spacecraft, maintaining an quasi-equilateral triangle and orbiting the Sun. 
Moreover, besides the waveform, GWs also retain polarization information, which can be used to test alternative theories of gravity with different polarization modes. 
In GR, GWs contain only the plus and cross tensor modes, while alternative theories can have up to four extra polarization modes~\cite{Eardley1973}. 
This paper focuses on the gravitational quantum field theory (GQFT)~\cite{Wu2016, Wu:2017urh, Wu2023, Wu:2024mul,Wu:2025abi,Wu:2025rei} which predicts five transverse polarization modes for the gravitational waves~\cite{Gao2024, Gao2025, Xu2025a}. 
Detection of non-tensorial polarizations will be direct evidence beyond GR, which is crucial for the development of gravitational theory.

In practical space-borne GW interferometers, the detector is composed of three spacecraft arranged in a quasi-equilateral triangular constellation~\cite{AmaroSeoane2017, Hu2017}, forming six one-way laser links. 
\begin{comment}
 Due to the dynamic motion of the constellation and unequal arm lengths, the actual measured signal is severely obscured by laser phase noise. 
Therefore, time-delay interferometry (TDI) are employed to introduce time shifts to individual link streams and combine them, thereby synthesizing a virtual equal-arm interferometer to suppress the dominant laser noise \cite{Tinto2021, Armstrong1999, Dhurandhar2002, Estabrook2000, Vallisneri2005}. 
Although TDI algorithms are highly effective in noise cancellation, they do not target specific polarization modes, and are particularly difficult to observe for transverse isotropic breathing mode.    
\end{comment}
From the six one-way data streams, we can construct the so-called null-response channel (NRC)~\cite{Tinto:2004nz, Barroso:2024azl, Xu2025} in which some polarizations can be highly suppressed. Building upon this, we further consider the response of extra polarization mode in tensor mode NRC (t-NRC), thereby providing a pathway for distinguishing alternative theories through polarization modes. Here we focus on testing GQFT in the t-NRC and verifying GR under the breathing mode NRC (b-NRC). Furthermore, we extend this scenario to dynamic constellation, making it more realistic for actual detection and enhancing the response of the breathing mode. We show that the NRC method can extract faint gravitational wave polarization signatures beyond GR, providing a powerful waveform independent approach for distinguishing different gravity theories. 

This paper is organized as follows. In Section.~\ref{Theoretical framework} we establish the theoretical framework by introducing briefly the formalism of GQFT and its corresponding polarization modes and revealing an extra breathing mode compared to GR.  Then we present the NRC method and construct null response for specific modes, enabling more effective tests of alternative theories. We further generalize to the dynamic orbits of LISA or Taiji, which enhances the response to the breathing mode. In Section.~\ref{Discriminate between GR and GQFT} we construct the t-NRC and the b-NRC to distinguish GQFT from GR, and present the corresponding polarization-averaged response functions and sensitivity curves. 
%Results are discussed for both the static constellation and the second-order orbit cases, and compared with the classic Michelson $X$ configuration. It is found that the NRC method provides a reliable criterion for distinguishing alternative theories of gravity. 
Finally, we give the conclusion in Section.~\ref{conclusion}. Throughout the texts, we use natural units ($\hbar=c=1$) unless specified otherwise.

\section{Theoretical framework \label{Theoretical framework}}

\subsection{GQFT and polarization modes}

GQFT offers a theoretical framework aimed at reconciling general relativity with quantum field theory~\cite{Wu2016, Wu:2017urh, Wu2023, Wu:2024mul,Wu:2025abi,Wu:2025rei}, thereby proposing an alternative theory of gravity that clearly distinguishes between the spacetime symmetry of coordinate systems and the intrinsic symmetry of the fundamental constituents of matter. Within this framework, the base manifold maintains global Poincaré symmetry PO(1,3), while locally exhibiting a Poincaré-type inhomogeneous spin gauge symmetry in the spinor representation of basic fermions. This local symmetry, denoted as $WS(1,3)=W^{1,3} \rtimes SP(1,3)$, governs the Weyl-type fermions in the Standard Model of elementary particles. Here, $W^{1,3}$ represents a translation-like chirality boost-spin gauge symmetry, while $SP(1,3)$ corresponds to a rotational-like spin gauge symmetry. 

The corresponding gauge fields are the chirality boost-spin gauge field $\mathcal{W}^a_\mu(x)$ and the spin gauge field $\mathcal{A}^{ab}_\mu(x)$. To preserve both the global Poincaré symmetry and the inhomogeneous spin gauge symmetry, the framework inevitably introduces a bi-covariant vector field $\hat{\chi}_{a}^{\mu}(x)$ and its dual $\chi^a_\mu (x)$. The field $\chi^a_\mu (x)$, referred to as the gravigauge field, behaves as a Goldstone-type boson identified to a massless graviton. The composite tensor field $\chi_{\mu\nu} = \eta_{ab} \chi^a_\mu \chi^b_\nu$ corresponds to the metric $g_{\mu\nu}$ of the spcetime manifold in GR. This demonstrates that the gravigauge field $\chi^a_\mu (x)$ provides a geometric bridge between the spin fiber gravigauge spacetime and the base spacetime manifold. 

The equation of motion for the gravigauge field leads to a gauge-type gravitational equation. When projected onto the coordinate base spacetime through the gravigauge field $\chi^a_\mu (x)$ and its dual $\hat{\chi}_{a}^{\mu}(x)$, GQFT yields the following equations~\cite{Wu:2024mul,Wu:2025abi,Wu:2025rei}:
\begin{gather}
    R_{\mu\nu} - \frac{1}{2} \chi_{\mu\nu} R + \gamma_{w} \mathcal{G}_{(\mu\nu)} = 8\pi G_{\kappa} T_{(\mu\nu)}, \label{symEq}\\
    \gamma_{w}\mathcal{G}_{[\mu\nu]} = 8\pi G_{\kappa} T_{[\mu\nu]},  \label{antiSymEq}.
\end{gather}
Here Eq. \eqref{symEq} is the symmetric equation, which leads to a generalized Einstein equation, and Eq. \eqref{antiSymEq} is an extra antisymmetric equation. 

Linearization of the GQFT dynamical equations reveals five polarizations~\cite{Gao2024}.
Among these are the transverse-traceless tensor $\hat{h}_{ij}$ in general relativity, two transverse vector modes $F_i$, and one transverse scalar mode $\Psi$.
The generation of these polarizations has been discussed in Ref.~\cite{Gao2025}, whereas current space-borne gravitational-wave detector such as LISA and Taiji rely solely on geometric observations and do not account for coupling effects between spin and gravity. And Ref.~\cite{Xu2025a} demonstrates through the geodesic deviation method to calculate the electric components of Riemann tensor, indicating that GQFT possesses three observable geometric polarizations, namely, two tensor modes (plus and cross as in GR), and an extra transverse scalar breathing mode. 

In this work, we focus on the response of three geometric polarization modes in LISA and Taiji. 
By constructing specific interferometric channels to distinguish GR and GQFT from a polarization perspective, we shall present the optimal observational combinations for the additional breathing mode, along with the corresponding responses and sensitivities. 
However, the other two transverse vector modes are entirely determined by the spin-related antisymmetric tensors and require further study once spin effect has been taken into account. 

\subsection{Signal Response and NRC }

We consider space-borne detectors with three spacecrafts arranged in a roughly equilateral triangle.
%, and utilizes Michelson laser interferometry to capture the gravitational-wave signal. 
Each pair of spacecrafts have optical benches that send and receive laser light from each other, resulting in six single-link data streams in total. The response of single-link to a GW propagating in the direction $\hat{k}$ can be expressed as \cite{Cornish2003,Cornish2003a}
\begin{equation}
    y_{ij}(t) =- \frac{\hat{r}_{ij} \otimes \hat{r}_{ij} : \mathbf{e}^p}{2 (1 - \hat{k} \cdot \hat{r}_{ij})} \left[ h_p(t - \hat{k} \cdot r_j) - h_p(t - L_{ij} - \hat{k} \cdot r_i) \right], \label{single link response}
\end{equation}
where ${r}_i$ and $r_j$ ($i,j=1,2,3$) are the position vectors of sending and receiving spacecrafts, respectively, and $\hat{r}_{ij}={(r_j-r_i)}/{L_{ij}}$ is the unit vector pointing from sending spacecraft to receiving spacecraft, with $L_{ij}=| r_j - r_i |$ corresponding to the arm length. 
$\mathbf{e}^p$ are the polarization tensors of GW, and $h_p(t)$ corresponds to the waveform in time domain of polarization $p$.
Here we adopt the ecliptic coordinate $\hat{k} = (- \cos \mathfrak{l} \cos \mathfrak{b}, - \cos \mathfrak{l} \sin \mathfrak{b}, - \sin \mathfrak{l}) \label{directionK}$. Details on single-link computation are provided in the Appendix~\ref{single-link computation}. 

However, the actual recorded single-link data are dominated by laser phase noise, we shall use time-delay interferometry (TDI) to time-shift and combine the response to form virtual equal-arm interferometers, aiming at effectively canceling the laser noise \cite{Tinto2021, Armstrong1999, Dhurandhar2002, Estabrook2000, Vallisneri2005}. Here and after, we shall focus on the first-generation channels as the sensitivities to GW are the same for the relevant second-generation ones~\cite{Tinto:2022zmf,Yu:2023iog}. For example, the standard Michelson channel $X$ in time domain is constructed as
\begin{equation}
    X(t) = (y_{13} + y_{31,2} + y_{12,22} + y_{21,322}) - (y_{12} + y_{21,3} + y_{13,33} + y_{31,233}), 
\end{equation}
where $y_{ij,mn\cdots}(t) = y_{ij}(t - L_m - L_n - \cdots)$, and $L_m$ denotes the length of the arm that is located opposite spacecraft $m$. 

%Here, it is assumed that the arm length is constant at the mean value of constellation. For a more general case, refer to the next subsection. \rem{better have a schematic triangle figure to show these $L_{m}$.}

Moreover, a generic configuration $\eta(t)$ can be represented as a linear combination of time-delayed Sagnac channels $\alpha$, $\beta$ and $\gamma$. Explicitly, we have 
\begin{equation}
    \alpha(t) = (y_{13} + y_{32,2} + y_{21,12}) - (y_{12} + y_{23,3} + y_{31,13}),
\end{equation}
The expressions for $\beta(t)$ and $\gamma(t)$ are obtained by applying the index cycle $(1\rightarrow 2 \rightarrow 3\rightarrow 1)$ to the formula above, once for $\beta$ and twice for $\gamma$. We can also do Fourier transformation and write the amplitude in frequency domain
\begin{equation}
    \tilde{\alpha}(f) = \sum_{p=+,\times,\cdots}\tilde{\alpha}_p(f,\hat{k})\tilde{h}_p(f).
\end{equation}
Here $\tilde{h}_p(f)$ is the GW waveform in Fourier domain for polarization $p$. Similarly, we can define the corresponding $\tilde{\beta}(f)$, $\tilde{\gamma}(f)$, $\tilde{\beta}_p(f,\hat{k})$ and $\tilde{\gamma}_p(f,\hat{k})$.

Then $\eta$ in Fourier domain can be described as 
\begin{equation} \label{NRC response}
    \tilde{\eta}(f) = a_1 \tilde{\alpha} + a_2 \tilde{\beta} + a_3 \tilde{\gamma} \equiv \mathbf{a}^T \cdot \mathbf{\tilde{A}} \cdot \mathbf{\tilde{h}}_p, 
\end{equation}
where $\mathbf{a} = (a_1, a_2, a_3)^T$ , $a_i$ are frequency-dependent coefficients, $\mathbf{\tilde{h}}_p = (\tilde{h}_+(f),\tilde{h}_\times(f),\cdots)^T$ and $\mathbf{\tilde{A}}$ denotes the matrix with elements $\tilde{\alpha}_p(f,\hat{k})$, $\tilde{\beta}_p(f,\hat{k})$ and $\tilde{\gamma}_p(f,\hat{k})$,
\begin{equation}
    \mathbf{\tilde{A}} = (\mathbf{\tilde{A}}_+, \mathbf{\tilde{A}}_\times,\cdots)^T, \; \mathbf{\tilde{A}}_p :=  (\tilde{\alpha}_p,\tilde{\beta}_p,\tilde{\gamma}_p)^T.
\end{equation}

The corresponding SNR of $\eta$ channel can be calculated as~\cite{Stanislav2021}
\begin{equation}
    \mathrm{SNR}_\eta^2 = 4 \int_0^{f_{max}} \dif f \frac{\lvert a_1 \tilde{\alpha} + a_2 \tilde{\beta} + a_3 \tilde{\gamma} \rvert^2}{\mathbf{a}^\dagger \mathbf{S}_N(f) \mathbf{a}} , \label{SNR}
\end{equation}
Note that the SNR is physically related purely to the real part. $\mathbf{S}_N(f)$ is a real positive-definite matrix to represent noise correlations across the Sagnac combinations, 
and specific elements may be included in the Appendix \ref{noise matrix}. 

The SNR formula suggests that it is possible to find a specific coefficient vector such that under this particular channel the SNR is null, which is known as the NRC \cite{Tinto:2004nz, Barroso:2024azl, Xu2025}. 
The NRC can be used to distinguish subtle differences between distinct signals, such as the breathing mode difference between the GR and GQFT in this paper. 
Since SNR is a positive-definite functional, it can only vanish if the numerator of integrand is identically zero for each polarization
\begin{equation}
    a_1 \tilde{\alpha}_p + a_2 \tilde{\beta}_p + a_3 \tilde{\gamma}_p = 0. 
\end{equation}
\begin{comment}
On the other hand, the response of the TDI combination can be decomposed into polarization modes $(\tilde{\alpha}, \tilde{\beta}, \tilde{\gamma})^T = \mathbf{\tilde{X}} \cdot \mathbf{\tilde{h}}_p$, 
where here denotes $\mathbf{\tilde{X}}$ as the response matrix of the Sagnac combinations to the polarization modes, 
This is a $3 \times p$ and $f,\hat{k}$ dependent matrix, $p$ is the number of polarization modes, and each row corresponds to one of the $\tilde{\alpha}$, $\tilde{\beta}$ and $\tilde{\gamma}$ response of GWs, 
and each column corresponds to all the Sagnac responses of each polarization $\mathbf{\tilde{x}}_p$, where $p$ can be at most $+, \times, x, y, b, l$. 
In other words, $\mathbf{\tilde{x}}_p$ is the polarization component of $\mathbf{\tilde{X}}$. 
Meanwhile, $\mathbf{\tilde{h}}_p$ is a waveform vector of different polarizations $\tilde{h}_p$ in the Fourier space. 
\end{comment}
Therefore, the NRC condition can be rewritten as
\begin{equation}\label{NRC condition}
     (\mathbf{\tilde{A}})^T \mathbf{a}^n = 0,
\end{equation} 
where superscript $n$ denotes those polarization modes associated with a null response. 
This is a system of linear equations designed to find the kernel of operator $\mathbf{\tilde{A}}$. 
It can be seen that if the number of polarization modes exceeds three, there is no NRC that can be null to all these polarizations. 
If $\mathbf{\tilde{A}}$ is full rank for three polarization modes, there is still no nontrivial solution, and we have checked that this is indeed the case for the GQFT. 
However, the extra breathing mode in GQFT is a special scalar mode that can not be interchanged with tensor modes plus and cross. 
Therefore, it is possible to consider tensor mode NRC (t-NRC) and breathing mode NRC (b-NRC) separately to distinguish between GR and GQFT. 
In the following discussion, we shall elaborate how to test GQFT by differentiating the tensor modes and breathing mode. 

For the t-NRC that is null with respect to the tensor modes, we can select the coefficient vector to be orthogonal to the plane spanned by the two vectors $\mathbf{\tilde{A}}_+$ and $\mathbf{\tilde{A}}_\times$.
\begin{equation}
    \mathbf{a}^{t}(f,\hat{k}) = \mathbf{\tilde{A}}_+ \times \mathbf{\tilde{A}}_\times = 
    \left(
    \begin{matrix}
\tilde{\beta}_+\tilde{\gamma}_\times - \tilde{\gamma}_+ \tilde{\beta}_\times \\
\tilde{\gamma}_+\tilde{\alpha}_\times - \tilde{\alpha}_+ \tilde{\gamma}_\times \\
\tilde{\alpha}_+\tilde{\beta}_\times - \tilde{\beta}_+ \tilde{\alpha}_\times 
\end{matrix}
    \right)
    . \label{coeffgrNRC}
\end{equation}
Similarly for the b-NRC, we can choose the coefficients 
\begin{equation}
    \mathbf{a}^b(f,\hat{k}) = (- \tilde{\beta}_b - \tilde{\gamma}_b,\tilde{\alpha}_b,\tilde{\alpha}_b)^T. \label{coeffGQFTNRC}
\end{equation}

Note that the derivation of the NRC coefficient Eq. \eqref{NRC condition} relies solely on the response of Sagnac combination. 
Consequently, the reliability of the NRC does not depend on any specific waveform $\tilde{h}_p(f)$. 
Once the NRC coefficients for specific modes have been fixed, the corresponding interferometric channel yields zero response for the modes, which can be used to identify the polarization of the signals. 

Explicitly, we can analytically evaluate the Sagnac response to monochromatic plane GW, with the $\tilde{\alpha}_p$ as an illustrative example
\begin{equation}\label{eq:alpha_p}
    \tilde{\alpha}_p = \sum_{i \ne j} C_{ij} F_{ij}^p,\; F_{ij}^p =  \dfrac{\hat{r}_{ij} \otimes \hat{r}_{ij} : \mathbf{e}^p}{2 (1 - \hat{k} \cdot \hat{r}_{ij})}, \quad i,j=1,2,3,
\end{equation}
%where $F_{ij}^p$ is denoted as $F_{ij}^p =  \dfrac{\hat{r}_{ij} \otimes \hat{r}_{ij} : \mathbf{e}^p}{2 (1 - \hat{k} \cdot \hat{r}_{ij})}$ with the polarization mode $\mathbf{e}^p$, also $\hat{k}$ representing the direction of GW propagation \ref{directionK}. 
as a combination of $F_{ij}^p$ and the corresponding coefficients are as detailed
\begin{equation} 
    \begin{aligned}
        C_{12} ={}& e^{- \mathrm{i} \omega (L_3 + \hat{k} \cdot r_1) } - e^{- \mathrm{i} \omega  \hat{k} \cdot r_2 },\\
        C_{13} ={}& e^{- \mathrm{i} \omega \hat{k} \cdot r_3 } - e^{- \mathrm{i} \omega  (L_2 + \hat{k} \cdot r_1) },\\
        C_{21} ={}& e^{- \mathrm{i} \omega (L_1 + L_2) } \bigl( e^{- \mathrm{i} \omega \hat{k} \cdot r_1 } - e^{- \mathrm{i} \omega  (L_3 + \hat{k} \cdot r_2) } \bigr),\\
        C_{23} ={}& e^{- \mathrm{i} \omega L_3 } \bigl( e^{- \mathrm{i} \omega  (L_1 + \hat{k} \cdot r_2) } - e^{- \mathrm{i} \omega  \hat{k} \cdot r_3 }\bigr),\\
        C_{31} ={}& e^{- \mathrm{i} \omega (L_1 + L_3) } \bigl( e^{- \mathrm{i} \omega (L_2 + \hat{k} \cdot r_3) } - e^{- \mathrm{i} \omega  \hat{k} \cdot r_1 }\bigr),\\
        C_{32} ={}& e^{- \mathrm{i} \omega L_2 } \bigl( e^{- \mathrm{i} \omega \hat{k} \cdot r_2 } - e^{- \mathrm{i} \omega (L_1 + \hat{k} \cdot r_3)}\bigr), 
    \end{aligned}    
\end{equation}
where $\omega := 2 \pi f$. The responses of the other two Sagnac combinations $\tilde{\beta}_p$ and $\tilde{\gamma}_p$ can be obtained through index permutation $(1\rightarrow 2 \rightarrow 3\rightarrow 1)$. Therefore we shall focus on the properties of $\tilde{\alpha}_p$ below.

In the low-frequency/long-wavelength region $2 \pi fL_{m} \ll 1$, we can expand the exponential factors and obtain the vanishing constant and linear terms on frequency $f$,
\begin{align}
\tilde{\alpha}_{p \, (0)} ={}& 0, \\
\tilde{\alpha}_{p \, (1)} ={}& \mathrm{i} 2 \pi f \bigl[ - F_{12}^p (L_3 + \hat{k} \cdot r_1 - \hat{k} \cdot r_2) +                        F_{21}^p (L_3 - \hat{k} \cdot r_1 + \hat{k} \cdot r_2) + F_{13}^p (L_2 + \hat{k} \cdot r_1 -                    \hat{k} \cdot r_3) \nonumber\\
                &- F_{23}^p (L_1 + \hat{k} \cdot r_2 - \hat{k} \cdot r_3) - F_{31}^p (L_2 - \hat{k} \cdot r_1 + \hat{k} \cdot r_3) + F_{32}^p (L_1 - \hat{k} \cdot r_2 + \hat{k} \cdot r_3) \bigr]\nonumber\\
                ={}& \mathrm{i} 2 \pi f \bigl[ - F_{12}^p (L_3 - L_3 \hat{k} \cdot \hat{r}_{12}) + F_{21}^p (L_3 - L_3 \hat{k} \cdot \hat{r}_{21}) + F_{13}^p (L_2 - L_2 \hat{k} \cdot \hat{r}_{13}) \nonumber\\
                & - F_{23}^p (L_1 - L_1 \hat{k} \cdot \hat{r}_{23})- F_{31}^p (L_2 - L_2 \hat{k} \cdot \hat{r}_{31}) - F_{32}^p (L_1 - L_1 \hat{k} \cdot \hat{r}_{32})\bigr] \equiv 0
\end{align}
Then it indicates that in such a case the GW response will be at least proportional to $f^2$ in low-frequency regime. We can calculate the corresponding term
\begin{equation}
    \begin{aligned}
        \tilde{\alpha}_{p \, (2)} 
\begin{comment}
         ={}& 2 \pi^2 f^2 \Bigl\{ F_{12}^p \bigl[ -(L_3 + \hat{k} \cdot r_1)^2 + (\hat{k} \cdot r_2)^2  \bigr] +F_{21}^p \bigl[- (\hat{k} \cdot r_1)^2 + (L_3 + \hat{k} \cdot r_2)^2 \\
        & - 2 (L_1 +L_2)(- L_3 +\hat{k} \cdot r_1 -\hat{k} \cdot r_2 )\bigr] + F_{13}^p \bigl[ (L_2 + \hat{k} \cdot r_1)^2 - (\hat{k} \cdot r_3)^2  \bigr] \\
        & +F_{23}^p \bigl[- (L_1 + \hat{k} \cdot r_2)^2  + ( \hat{k} \cdot r_3)^2 - 2 L_3 ( L_1 +\hat{k} \cdot r_2 -\hat{k} \cdot r_3) \bigr]   \\
        & +F_{32}^p \bigl[-( \hat{k} \cdot r_2 )^2- 2 L_2 ( - L_1 +\hat{k} \cdot r_2 -\hat{k} \cdot r_3) +  (L_1 + \hat{k} \cdot r_3)^2 \bigr] \\
        &+F_{31}^p \bigl[( \hat{k} \cdot r_1 )^2 + 2 (L_1 +L_3) ( - L_2 +\hat{k} \cdot r_1 -\hat{k} \cdot r_3) -  (L_2 + \hat{k} \cdot r_3)^2 \bigr]\Bigr\}\\
       ={}& 2 \pi^2 f^2 \bigl[ 2 (L_1 +L_2) L_3 F_{21}^p (1 -\hat{k} \cdot \hat{r}_{21}) -2 L_3 L_1 F_{23}^p (1 -\hat{k} \cdot \hat{r}_{23})  \\
        &+2 L_2 L_1 F_{32}^p (1 -\hat{k} \cdot \hat{r}_{32})- 2 (L_1 +L_3) L_2 F_{31}^p (1 -\hat{k} \cdot \hat{r}_{31})\bigr] \\   
\end{comment}        
        = 4 \pi^2 f^2 \Bigl\{ & L_1 L_2 \bigl[ (F_{32}^p (1 -\hat{k} \cdot \hat{r}_{32}) - F_{31}^p (1 -\hat{k} \cdot \hat{r}_{31}) \bigr]  \\
        & + L_1 L_3 \bigl[ (F_{21}^p (1 -\hat{k} \cdot \hat{r}_{21}) - F_{23}^p (1 -\hat{k} \cdot \hat{r}_{23}) \bigr] \\
        &+ L_2 L_3 \bigl[ (F_{21}^p (1 -\hat{k} \cdot \hat{r}_{21}) - F_{31}^p (1 -\hat{k} \cdot \hat{r}_{31}) \bigr]  \Bigr\} \label{alpha2}.
    \end{aligned}
\end{equation}
We did not list terms on next orders, as we are using the full form of Eq.~\ref{eq:alpha_p} in the actual numerical calculation. However, the preceding discussion already suggests, at least in part, that the response function will exhibit a power-law dependence on $f$ in the low-frequency regime, provided that a contribution of a particular order is dominant.
%Where the quadratic terms can cancel each other out, taking index 12 as an example, the corresponding term $L_3 \mleft[ L_3 + \hat{k} \cdot (r_1 + r_2) \mright] \mleft[ F_{21}^p (1 -\hat{k} \cdot \hat{r}_{21} ) - F_{12}^p (1 -\hat{k} \cdot \hat{r}_{12} ) \mright]  \equiv 0 $, due to the symmetry of $F_{ij}^p (1 -\hat{k} \cdot \hat{r}_{ij})$, the other two terms can be similarly obtained. Thus, first order term still dominates the response, 

\subsection{Dynamic Constellation \label{orbit effect}}

The above discussion has considered the static scenario in the spacecraft coordinate system, which assumes a fixed angle between the monochromatic source and the spacecraft, thereby facilitating the calculation of the cumulative response to GWs. However, both LISA and Taiji actually orbit the Sun in the barycenter framework of the solar system. Therefore, in addition to the conventional NRC comparison between GR and GQFT, we further consider scenarios involving orbital effects in the barycenter frame. 

In this heliocentric ecliptic coordinate system, the z-axis is aligned with the orbital angular momentum of Earth, while the x-axis points to the vernal equinox, and a right-handed coordinate system is established on the basis of this. 
Hereby, orbital trajectories for spacecraft under second-order eccentricity can be determined \cite{Rubbo2004}
\begin{align}
    x_i(t) = & R \cos \alpha_t + \frac{1}{2} e R \left[\cos (2\alpha_t - \beta_i) - 3 \cos \beta_i\right] \nonumber \\ 
    &+ \frac{1}{8} e^2 R \left[3 \cos (3 \alpha_t - 2 \beta_i) - 10 \cos \alpha_t - 5 \cos (\alpha_t - 2 \beta_i)\right], \\
    y_i(t) = & R \sin \alpha_t + \frac{1}{2} e R \left[\sin (2\alpha_t - \beta_i) - 3 \sin \beta_i\right] \nonumber \\ 
    &+ \frac{1}{8} e^2 R \left[3 \sin (3 \alpha_t - 2 \beta_i) - 10 \sin \alpha_t + 5 \sin (\alpha_t - 2 \beta_i)\right], \\
    z_i(t) = & - \sqrt{3} e R \cos (\alpha_t - \beta_i) + \sqrt{3} e^2 R \left[\cos^2 (\alpha_t - \beta_i) + 2 \sin^2 (\alpha_t - \beta_i)\right]. \label{orbit}
\end{align}
Where $R = 1 \, \mathrm{AU}$ denotes the radial distance to the guiding center, $L=3\times10^9$~m is the fiducial arm length of the constellation, and $e = \frac{L}{2 \sqrt{3} R}\simeq 0.006$ is the eccentricity of the orbit. 
$\alpha_t = 2 \pi f_m t + \kappa$ (the modulation frequency $f_m = 1 / \mathrm{year}$) is the orbital phase and encodes the time dependence of the spacecraft's position. $\beta_i = \frac{2 \pi i}{3} + \lambda \, (i = 1,2,3)$ describes the relative phases of three spacecrafts within the constellation. 
Here and after we set initial ecliptic longitude and orientation of the constellation to zero, $\kappa = \lambda = 0$.

With the orbital trajectories of three spacecraft, the corresponding arm lengths can be obtained
\begin{align}
    L_{12}(t) &= 2 \sqrt{3} eR \left\{1 + \frac{e}{32} \left[ 15 \sin \left(\alpha_t - \lambda + \frac{\pi}{6}\right) - \cos 3(\alpha_t - \lambda)\right]\right\}, \\
    L_{13}(t) &= 2 \sqrt{3} eR \left\{1 - \frac{e}{32} \left[ 15 \sin \left(\alpha_t - \lambda - \frac{\pi}{6}\right) + \cos 3(\alpha_t - \lambda)\right]\right\}, \\
    L_{23}(t) &= 2 \sqrt{3} eR \left\{1 - \frac{e}{32} \left[ 15 \cos (\alpha_t - \lambda) + \cos 3(\alpha_t - \lambda)\right]\right\}.
\end{align}
Note that arm lengths are constant on leading order of $e$. However, for the uniform and weak mode, such as the extra breathing mode in GQFT, the mode is more sensitive in the low-frequency region to variations in arm length. In such a case, it is necessary to account for dynamic changes of the triangular constellation. Therefore, the calculation starting from single-link response Eq.~\eqref{single link response} takes into account the time dependence of both the spacecraft’s orbit and the relative motion. 

\begin{figure}[!htbp]
    \centering
            \includegraphics[width=0.55\textwidth]{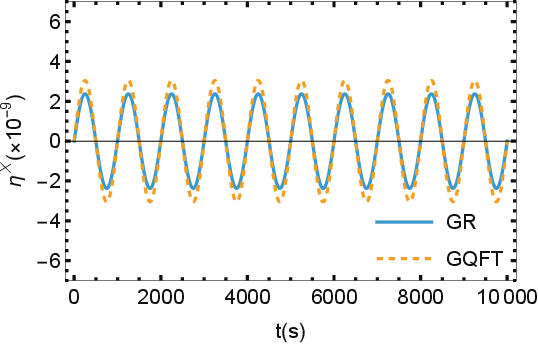}
            \includegraphics[width=0.55\textwidth]{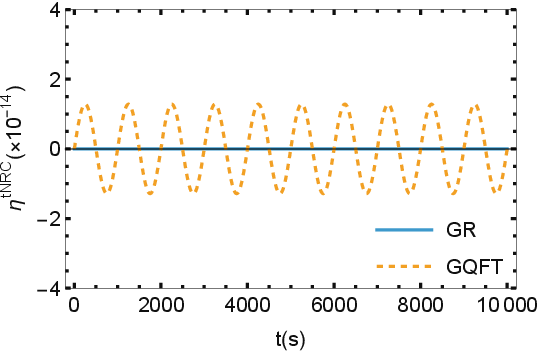}
            \includegraphics[width=0.55\textwidth]{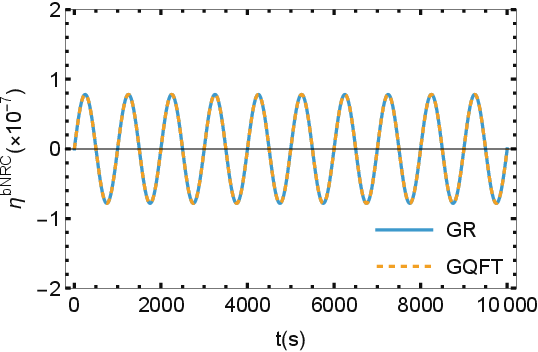}
    \caption{Illustration of the signals in Michelson $X$ and NRC in the time domain for unit amplitude and $f = 1 \, \mathrm{mHz}$ GWs incorporating different polarization modes in GR and GQFT, and the direction of NRC configuration is $(\mathfrak{l},\mathfrak{b}) = (0.88,1.77)$. 
            Across all figures, the GR and GQFT signals are respectively denoted by solid blue and dashed orange curves. 
            The top figure shows the signals in $X$. The middle one shows the signals in t-NRC, which indicates that the tensor mode is completely suppressed while only a faint signal from the breathing mode remains. The bottom figure shows the signals in b-NRC and the suppression of the breathing mode renders tensor signals identical in both theories. } \label{sigTime}
\end{figure}

To provide a visual illustration, we show in Fig.~\ref{sigTime} the responses of Michelson $X$ channel and NRC in time domain to monochromatic GWs with sky position $(\mathfrak{l},\mathfrak{b}) = (0.88,1.77)$.  It can be seen that under t-NRC, the tensor mode is effectively suppressed, whilst the breathing mode is highlighted. However, the breathing mode also appears to be moderately affected, which stems from the fact that both the breathing mode and the tensor modes are transverse and breathing mode is a uniformly isotropic mode. 
From the perspective of detection response, TDI operations inevitably partially cancel the breathing mode strength. 
Physically speaking, the breathing mode as a scalar mode is fundamentally different from tensor mode. Although the signal is weaker than normal signal, the breathing mode response can still be filtered out under t-NRC. 
This further illustrates that detecting the breathing mode in the time domain is difficult; indeed, the distinction may be more pronounced in the frequency domain. And, as expected that under b-NRC, the responses of GR and GQFT are completely identical, this can be regarded as a form of validation of tensor modes from the GQFT perspective. 

\section{Discriminate between GR and GQFT \label{Discriminate between GR and GQFT}}

This section will employ the NRC configuration to analyze monochromatic plane GW signals, with the aim of distinguishing between GR and GQFT via tensor mode NRC (t-NRC) and breathing mode NRC (b-NRC). Note that t-NRC can yield non-vanishing signals in GQFT,  while b-NRC can provide confirmation of GR. 
Furthermore, comparing their sensitivity curves with that of standard Michelson $X$ channel demonstrates the usefulness of the NRC method. 
%For convenience in this paper, NRC index will be denoted with superscript, and observed index with subscript. 

Starting with Eqs.~\eqref{NRC response}, \eqref{coeffgrNRC} and \eqref{coeffGQFTNRC}, we denote the response function for the corresponding polarization as
\begin{equation}
    R^n_{p}(f,\hat{k}) = \mathbf{a}^n \cdot \mathbf{\tilde{A}}_p,
\end{equation}
where the superscript $n$ indicates the mode response suppressed by NRC, the subscript $p$ corresponds to the signal to be observed. 
The polarization-averaged response functions of tensor mode and breathing mode on each other's NRC are provided
\begin{comment}
\begin{equation}
    R^n_{s}(f,\hat{k}) = \sqrt{\sum_{p=1}^P \frac{\mleft\lvert R^n_{s,p}(f,\hat{k}) \mright\rvert^2}{P}}, 
\end{equation}
\end{comment}
\begin{equation}
    R^{b}(f,\hat{k}) = \sqrt{{\left[| R^b_{+}(f,\hat{k}) |^2+| R^b_{\times}(f,\hat{k})|^2\right]}/{2}}, \; R^{t}(f,\hat{k}) = | R^t_{b}(f,\hat{k})|.
\end{equation}
The above quantities describe how tensor (breathing) modes response in b-NRC (t-NRC).
Without loss of generality, later $\hat{k}$ is selected to three sky positions $(\mathfrak{l},\mathfrak{b}) = (0.88,1.77)$, $(0.80,0.05)$ and $(0.05,1.77)$.

Once the corresponding GW response functions have been calculated, they can be compared with those of $X$ configuration to demonstrate the effectiveness of the NRC method. 
Therefore, a further comparison of the sensitivity curves is required, still continuing from the SNR Eq. \eqref{SNR}
\begin{equation}
    \mathrm{SNR}_\eta^2 = 4 \int_0^{f_{max}} \dif f \frac{\tilde{\eta}(f) \tilde{\eta}^*(f)}{S_n(f)} , 
\end{equation}
where $S_n(f):=\mathbf{a}^{n\dagger} \mathbf{S}^N(f) \mathbf{a}^n$ is the one-sided noise PSD and is calculated following Eq. \eqref{noiseX}. 
Additionally, since the signal $\tilde{\eta}(f)$ can be expressed as $\tilde{\eta}(f) = \sum_p R^n_{p}(f,\hat{k}) \tilde{h}_p $, substituting $\tilde{\eta}(f)$ in the SNR and averaging over the sky and polarizations, we can rewrite
\begin{equation}
    \mleft\langle \mathrm{SNR}_\eta^2 \mright\rangle = 4 \int_0^{f_{max}} \dif f \frac{(R^n(f))^2 \mathbf{\tilde{h}}_p \mathbf{\tilde{h}}_p^*}{S_n(f)} =4  \int_0^{f_{max}} \dif f \frac{\mathbf{\tilde{h}}_p \mathbf{\tilde{h}}_p^*}{S_h(f)} .
\end{equation}
Thus, the sensitivity denotes $S_h(f) := \dfrac{S_n(f)}{(R^n(f))^2}$, and $\sqrt{S_h(f)}$ will be used later in this paper. 
Specifically, PSD provides a quantitative framework for defining the noise floor, which enables the determination of the SNR for various GW sources. 
In a sense, the characterization of the PSD implies observational predictability. For convenience, we analyze the PSD curve under a unit SNR condition and provide the ratio of the SNR for the NRC configuration to Michelson $X$ to facilitate a clear comparison. 

\subsection{Response and Sensitivity in a Static Constellation.}
\begin{figure}[!htbp]
    \centering
            \includegraphics[width=0.49\textwidth]{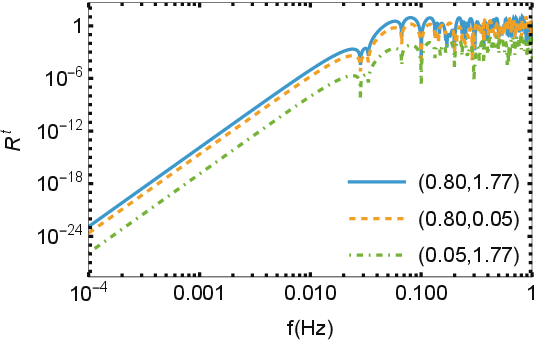}
            \includegraphics[width=0.49\textwidth]{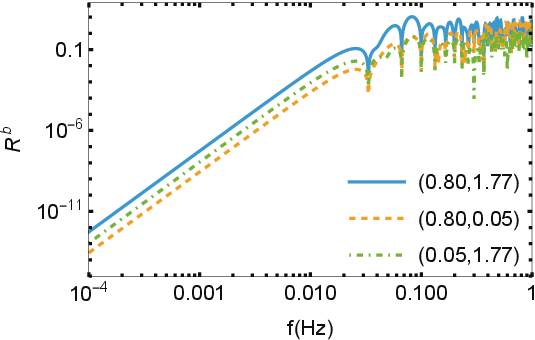}
    \caption{The response functions for a static constellation. The left figure shows the response function of breathing mode in t-NRC, and the right one shows response function of tensor mode in b-NRC.  } \label{responseStatic}
\end{figure}
This subsection presents the signals from monochromatic plane GWs observed, respectively, under tensor mode and breathing mode in each other NRC with a static constellation. The coordinate here is chosen on x-y plane, with one spacecraft as the origin and the x-axis aligned with the median of constellation triangle, namely $(0,0,0)$, $L(\cos \frac{\pi}{6}, - \sin \frac{\pi}{6},0)$ and $L(\cos \frac{\pi}{6}, \sin \frac{\pi}{6},0)$, where $L$ is the arm length. 
%Subsequently, following the above process, we show the corresponding response functions. 

Fig. \ref{responseStatic} illustrates that the response functions are linear in the low-frequency regime on a log–log plot, indicating a power-law behavior in the relation between the response function and frequency. We find that in low-frequency regime response function of the breathing mode in t-NRC $R^{t} \propto f^9$, and response function of the tensor mode in b-NRC $R^{b} \propto f^5$. 
Compared to the distinction between GWs and ultralight dark matter, which is characterized by a two-regime frequency dependence also within the static frame \cite{Xu2025}, the distinction between the tensor modes and breathing mode of GWs follows an identical power-law at low frequencies and is more challenging to distinguish in intensity. 
Therefore, the extra breathing mode response of GQFT is not as strong as the tensor mode of GR. 
This can be explained by the fact that the breathing mode, as an isotropic mode, shows little difference in its projection across each arm, which imposes higher order of detection. 

\begin{figure}[!htbp]
    \centering
            \includegraphics[width=0.7\textwidth]{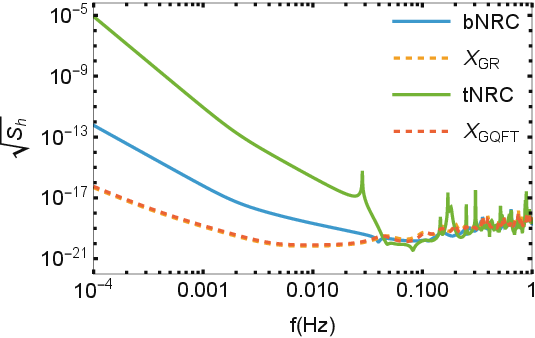}
            \includegraphics[width=0.7\textwidth]{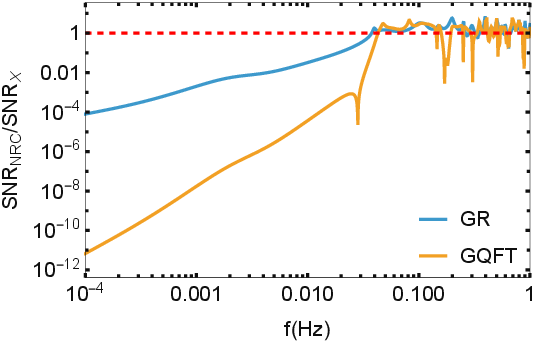}
    \caption{The top figure shows sensitivity curves for different configurations: the cyan solid line corresponds to tensor mode in the b-NRC marked GR, the orange dashed line corresponds to GR in the Michelson-like X, the green solid line corresponds to breathing mode in the t-NRC marked GQFT, and the vermilion dashed line corresponds to GQFT in the Michelson-like X. 
    The bottom figure shows the ratio of the SNR for NRC and the Michelson $X$ configurations under the same theory: the blue line represents the detection of GR, the orange line represents the detection of GQFT, and the red dashed line represents a ratio of 1. } \label{senStatic}
\end{figure}
Then in Fig.~\ref{senStatic} we show the sensitivity curves for different NRCs and Michelson $X$, along with the ratio of their SNRs. The figure can be understood as follows. Once a strong GW signal has been identified in Michelson $X$ channel, we can further calculate the SNRs in t-NRC and b-NRC, and their ratio to $X$ channel. If there is a breathing mode in the signal, then the ratio should be aligned with the orange curve in the bottom figure. For tensor modes the ratio should follow the blue curve. 
If used alone, the NRC configuration would be less sensitive in the low-frequency regime compared to the Michelson $X$ channel. 
However, in the high-frequency band approaching $0.05 - 1 \, \mathrm{Hz}$, the NRC method can distinguish different GW polarization modes without sensitivity loss in general.

\begin{comment}
It illustrates that the NRC method can detect non-zero responses from distinct modes.
%specifically, the breathing mode exhibits a spike at $0.028 \, \mathrm{Hz}$, which may be a signature of scalar modes. 
Meanwhile their sensitivities are degraded in the low-frequency regime compared to the Michelson $X$.
%the observation of extra breathing mode is more challenging for tensor modes. 
This situation arises due to the NRC method does not merely seek to maximize signal strength, but rather precisely suppresses unneeded modes to highlight the desired signal features. 
Therefore, while the NRC configuration trades off some sensitivity in the low-frequency regime in order to improve the target signal purity. 
This provides a better basis for resolving modes that are difficult to analyze using conventional methods, and further demonstrating that the NRC method is effective in distinguish between GR and GQFT from a polarization perspective.     
\end{comment}

\subsection{Response and Sensitivity in a Dynamic Constellation}
\begin{figure}[!htbp]
    \centering
            \includegraphics[width=0.49\textwidth]{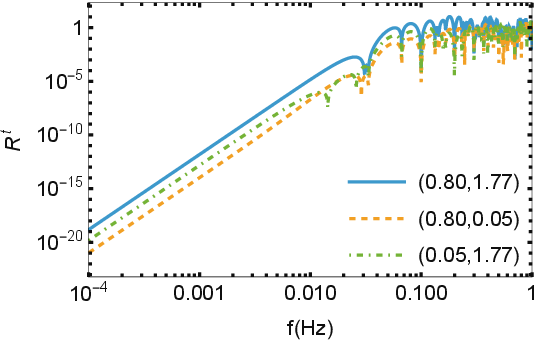}
            \includegraphics[width=0.49\textwidth]{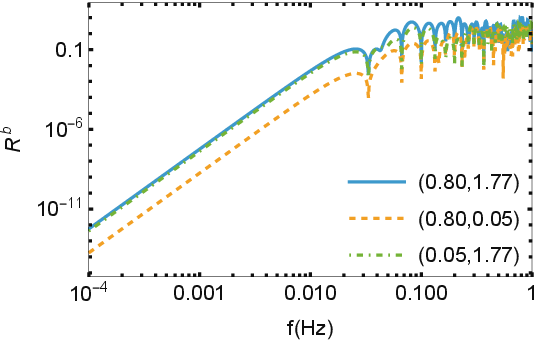}
    \caption{The response functions for a dynamic constellation. The left figure shows the response function of breathing mode in t-NRC, and the right one shows response function of tensor mode in b-NRC. } \label{responseOrbit}
\end{figure}

This subsection will generalize the above results to the case of more realistic orbits, with the constellation following Eq. \eqref{orbit}. 
Fig. \ref{responseOrbit} demonstrates that the response functions can have different behaviors in the low-frequency regime, namely, $R^{t} \propto f^7$ and $R^{b} \propto f^5$. 
It can be seen that the response to breathing mode under t-NRC is significantly improved by orbital effects, and this improvement is greater than the impact on tensor mode, whose response for a dynamic constellation may exhibit a variation of 20\% compared to static condition, even the effects are more pronounced in the high-frequency regions~\cite{shi2026construction}.
%It is evident that the response functions with orbital effects are significantly enhanced in the low-frequency regime for the breathing mode in t-NRC. 
%Also due to breathing mode is isotropic, the orbital effects introduce time-varying arm lengths and relative motion between spacecrafts, which may modulate the GW signal and enhance the response. 
Therefore, compared to tensor modes, orbital effects can significantly improve the response of the breathing mode. 

%Then, exhibit the sensitivity curves with orbital effects, along with the SNRs of NRC and the Michelson-like X.  
\begin{figure}[!htbp]
    \centering
            \includegraphics[width=0.7\textwidth]{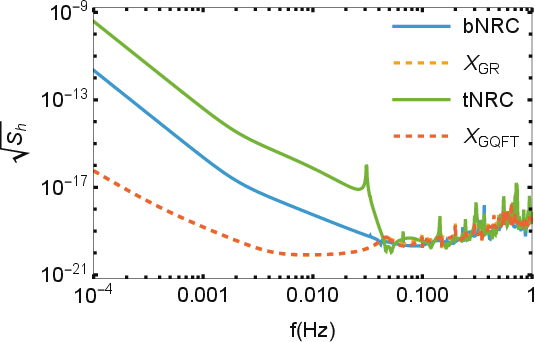}
            \includegraphics[width=0.7\textwidth]{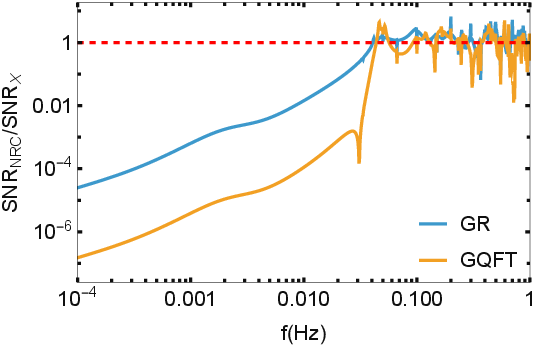}
    \caption{The top figure shows the sensitivity curves for different configurations under the second-order SSB orbits, with the same label as in Fig. \ref{senStatic}. 
    The bottom figure shows the ratio of NRC to Michelson $X$ with the orbital effects.} \label{senOrbit}
\end{figure}
Fig. \ref{senOrbit} illustrates the sensitivity curves in the dynamic case, similar to the static case in Fig.~\ref{senStatic}. 
%, but the sensitivity is significantly improved compared to the static case. 
However, we highlight that the sensitivity for breathing mode in t-NRC is enhanced by four orders of magnitude in low-frequency regime. 
The reason is that the response of breathing mode is enhanced with the isotropy breaking for a dynamic constellation. 
%Indeed, even in the first-order arm lengths are constant, then the breathing mode is not particularly pronounced, thus it is necessary to consider second-order orbits. 
This further demonstrates the importance of considering orbital dynamics when detecting GWs, and in this context, the NRC method is essential for distinguishing between different theories. 

\begin{comment}
Overall, although the extra breathing mode in GQFT is more challenging than tensor modes to detect, the NRC method can effectively suppress specific modes to enhance the detectability of other modes, making it more feasible to distinguish between GR and GQFT. 
Meanwhile, the t-NRC configuration is more suitable for detecting the breathing mode to validate GQFT, while the b-NRC configuration is more suitable for verifying GR, especially considering orbital effects.    
\end{comment}

\section{Conclusion \label{conclusion}}

We have described a method for testing GQFT at LISA and Taiji by its extra breathing scalar polarization. We can leverage the various interferometric channels and construct the null-response channel (NRC) with respect to some specific polarization or mode. For instance, the tensor NRC (t-NRC) still has response to the breathing mode. Therefore, a signal appearing in the t-NRC would indicate alternative theories of gravity. 

We have computed the response functions and sensitivity curves for various NRC and compared them with those in Michelson channel. The SNR ratio between different channels can provide a quantitative way to identify the polarization of the signal. Furthermore, we have taken into account the effect of dynamic constellation and found the response of breathing mode in t-NRC is enhanced by four orders of magnitude in low-frequency regime. This demonstrates that orbital effects should be incorporated in realistic data analysis. 
Meanwhile, in the high-frequency regime, different polarization modes are effectively distinguished by the NRC method without sensitivity loss. 
This study establishes a systematic procedure to distinguish the polarization modes of GWs, thereby providing a valuable methodology to probe alternative theories of gravity in future space-borne GW missions \cite{lin2024limiting,liang2026probing,wu2024comparison,gao2023testing,zhou2025sensitivity,lu2026gravitational}.

\begin{acknowledgments}
This work is partly supported by the National Key Research and Development Program of China (Grant No.~2021YFC2201901), the National Natural Science Foundation of China (NSFC) under Grants Nos. 12547104, 12441504, 12147103, the Strategic Priority Research Program of the Chinese Academy of Sciences and the Fundamental Research Funds for the Central Universities. 
\end{acknowledgments}

\appendix

\section{Single-link response \label{single-link computation}}

Considering the monochromatic plane GW response of a single-link, the espression can be written as
\begin{equation}
    y_{ij}(t) = \frac{\hat{r}_{ij} \otimes \hat{r}_{ij} : \mathbf{e}^p}{2 (1 - \hat{k} \cdot \hat{r}_{ij})} \mleft[ e^{- \mathrm{i} \omega \hat{k} \cdot r_j} - e^{- \mathrm{i} \omega ( L_{ij} + \hat{k} \cdot r_i)}  \mright] e^{\mathrm{i} \omega t} \label{y_ij},
\end{equation}
where $\mathbf{e}^p$ represents the polarization modes of GW, there can be up to six polarizations $+, \times, x, y, b, l$, including tensor, vector and scalar modes. 
In GR, there are only two tensor modes $+$ and $\times$, whereas in GQFT there is an extra scalar mode $b$. 
\begin{equation}
    \begin{gathered}
        \mathbf{e}^+ = \hat{u} \otimes \hat{u} - \hat{v} \otimes \hat{v}, \\
        \mathbf{e}^\times = \hat{u} \otimes \hat{v} + \hat{v} \otimes \hat{u}, \\
        \mathbf{e}^b = \hat{u} \otimes \hat{u} + \hat{v} \otimes \hat{v}. 
    \end{gathered}
\end{equation}
Meanwhile, a right-handed coordinate system based on unit vectors $\hat{k}$, $\hat{u}$ and $\hat{v}$ in the ecliptic coordinate system is adopted here, with $\hat{k}$ representing the direction of GW propagation \ref{directionK}. 
$\hat{u}$ and $\hat{v}$ are defined as follows
\begin{equation}
    \begin{gathered}
        \hat{u} = (- \sin \mathfrak{l} \cos \mathfrak{b}, - \sin \mathfrak{l} \sin \mathfrak{b}, \cos \mathfrak{l}), \;
        \hat{v} = (- \sin \mathfrak{b}, \cos \mathfrak{b}, 0). 
    \end{gathered}
\end{equation}

In the static case in which orbital evolution is neglected, we can get the response in Fourier domain
\begin{equation}
    \tilde{y}_{ij}(f) = \frac{\hat{r}_{ij} \otimes \hat{r}_{ij} : \mathbf{e}^p}{2 (1 - \hat{k} \cdot \hat{r}_{ij})} \mleft[ e^{- \mathrm{i} 2 \pi f \hat{k} \cdot r_j} - e^{- \mathrm{i} 2 \pi f ( L_{ij} + \hat{k} \cdot r_i)}  \mright] \delta(f - f_{gw}). 
\end{equation}

\section{Noise Matrix of Sagnac Combinations \label{noise matrix}}

The noise correlation matrix of Sagnac channels is as follows
\begin{equation}
    \mathbf{S}^N(f) = 
    \begin{bmatrix}
        S_{\alpha\alpha} & S_{\alpha\beta} & S_{\alpha\gamma} \\
        S_{\beta\alpha} & S_{\beta\beta} & S_{\beta\gamma} \\
        S_{\gamma\alpha} & S_{\gamma\beta} & S_{\gamma\gamma}
    \end{bmatrix},
\end{equation}
where the diagonal elements are the noise power spectral density (PSD) of each Sagnac channel $\alpha$, $\beta$ and $\gamma$, and the off-diagonal elements denote the cross-correlation PSD between different Sagnac combinations. The expressions of these elements can be calculated as \cite{Hartwig2023}
\begin{gather}
    S_{\alpha\alpha} = S_{\beta\beta} = S_{\gamma\gamma} = 6 S_{oms} + 4 (3 - 2\cos \delta - \cos 3\delta) S_{acc}, \\
    S_{\alpha\beta} = S_{\alpha\gamma} = S_{\beta\gamma} = 2 (2 \cos \delta + \cos 2\delta) S_{oms} - 4 (1 - \cos \delta) S_{acc}. 
\end{gather}
Here $\delta = 2 \pi f L$, and $L$ is the fiducial arm length of the triangle constellation, $L = 3 \times 10^9 \, \mathrm{m}$ for Taiji.
In the case of Michelson $X$ channel, the noise can be specified directly \cite{Yu:2023iog}
\begin{equation}
    S_X(f) = 16 \sin^2 \delta \mleft[ S_{oms} + (3 + \cos 2\delta) S_{acc} \mright]. \label{noiseX}
\end{equation}
Above, $S_{oms}$ and $S_{acc}$ are the optical metrology system noise and the acceleration noise, respectively. 
Their parametrization can be expressed as \cite{Stanislav2021}
\begin{gather}
    S_{oms} = \mleft(\frac{2 \pi f s_{oms}}{c}\mright)^2 \mleft[1 + \mleft(\frac{2 \times 10^{-3} \, \mathrm{Hz}}{f}\mright)^4\mright] \, \mathrm{Hz}^{-1}, \\
    S_{acc} = \mleft(\frac{s_{acc}}{2 \pi f c}\mright)^2 \mleft[1 + \mleft(\frac{0.4 \times 10^{-3} \, \mathrm{Hz}}{f}\mright)^2\mright] \mleft[1 + \mleft(\frac{f}{8 \times 10^{-3} \, \mathrm{Hz}}\mright)^4\mright] \, \mathrm{Hz}^{-1}. 
\end{gather}
For Taiji the parameters are $s_{oms} = 8 \times 10^{-12} \, \mathrm{m}$ and $s_{acc} = 3 \times 10^{-15} \, \mathrm{m/s^2}$ \cite{Xu2025}. 

\bibliography{distGRnGQFT}

\end{document}

%% file: cassX.tex
\usepackage[dvipsnames]{xcolor} %提供对颜色的支持，颜色包放最前，防止颜色表加载的bug，可选项里就是一张颜色表。

\usepackage{amsmath} %一个数学工具包
\usepackage{amsfonts} %提供数学字体
\usepackage{amsthm} %提供定理环境
\usepackage{amssymb} %提供数学符号，\mathfrak{…}, \mathbb{…}等
\usepackage{eucal} %把 \mathcal{…} 命令变得好看一些
\usepackage{mathrsfs} %提供 \mathscr{…} 命令
\usepackage{mathtools} %数学扩展包
\usepackage{array} %表格之类的
\usepackage{latexsym} %提供符号
\usepackage{slashed} %提供 \slashed{} 命令，在对应字符面上加上斜线
\usepackage{graphicx} %支持插入图片
\usepackage{mleftright} %提供 \mleft 和 \mright 命令(相当于优化的 \left \right命令)
\usepackage{enumitem} %可用 enumitem 宏包定制各种列表间距，还提供了对列表标签、引用等的定制。具体可看相应帮助文档。
\usepackage{tabularx} %提供一个 X 列格式，类似 p 列格式，不过会根据表格宽度自动计算列宽，多个 X 列格式平均分配列宽。
\usepackage{multirow,makecell} 
\usepackage{float} %为浮动体提供了 H 位置参数，不与 htbp 及 ! 混用。使用 H 位置参数时，会取消浮动机制，将浮动体视为一般的盒子插入当前位置。
\usepackage{caption,subcaption} % caption 提供了\caption*{}, 控制浮动体标题的格式， subcaption 提供子图表和子标题的排版。
\usepackage{verbatim} %优化了 verbatim 环境的内部命令，并提供了 \verbatiminput 命令用来直接读入文件生成代码环境。

\usepackage{textgreek}%提供了文本的希腊字母，\text+对应字母，标题文本用，且可规避一些不支持希腊字母的字体

\usepackage{makeidx} %调用 makeidx 宏包，用来处理索引
\makeindex %开启索引的收集

\usepackage[bookmarks=true,bookmarksopen=true, bookmarksnumbered=true]{hyperref}  %超链接宏包放在最后
\hypersetup{colorlinks = true} %设置hyperref宏包，为超链接文字带上颜色

\usepackage{orcidlink}%提供 ORCiD 插入，如\author{Emmy Noether\,\orcidlink{0000-0000-0000-0000}}, 注意名字和ORCiD的缩进。该宏包调用了 xcolor 和 hyperref 要放在它们后面。

\newcommand*{\dif}{\mathop{}\!\mathrm{d}} %定义普通微分算子d

%% file: main.bbl
\begin{thebibliography}{37}%
\makeatletter
\providecommand \@ifxundefined [1]{%
 \@ifx{#1\undefined}
}%
\providecommand \@ifnum [1]{%
 \ifnum #1\expandafter \@firstoftwo
 \else \expandafter \@secondoftwo
 \fi
}%
\providecommand \@ifx [1]{%
 \ifx #1\expandafter \@firstoftwo
 \else \expandafter \@secondoftwo
 \fi
}%
\providecommand \natexlab [1]{#1}%
\providecommand \enquote  [1]{``#1''}%
\providecommand \bibnamefont  [1]{#1}%
\providecommand \bibfnamefont [1]{#1}%
\providecommand \citenamefont [1]{#1}%
\providecommand \href@noop [0]{\@secondoftwo}%
\providecommand \href [0]{\begingroup \@sanitize@url \@href}%
\providecommand \@href[1]{\@@startlink{#1}\@@href}%
\providecommand \@@href[1]{\endgroup#1\@@endlink}%
\providecommand \@sanitize@url [0]{\catcode `\\12\catcode `\$12\catcode
  `\&12\catcode `\#12\catcode `\^12\catcode `\_12\catcode `\%12\relax}%
\providecommand \@@startlink[1]{}%
\providecommand \@@endlink[0]{}%
\providecommand \url  [0]{\begingroup\@sanitize@url \@url }%
\providecommand \@url [1]{\endgroup\@href {#1}{\urlprefix }}%
\providecommand \urlprefix  [0]{URL }%
\providecommand \Eprint [0]{\href }%
\providecommand \doibase [0]{https://doi.org/}%
\providecommand \selectlanguage [0]{\@gobble}%
\providecommand \bibinfo  [0]{\@secondoftwo}%
\providecommand \bibfield  [0]{\@secondoftwo}%
\providecommand \translation [1]{[#1]}%
\providecommand \BibitemOpen [0]{}%
\providecommand \bibitemStop [0]{}%
\providecommand \bibitemNoStop [0]{.\EOS\space}%
\providecommand \EOS [0]{\spacefactor3000\relax}%
\providecommand \BibitemShut  [1]{\csname bibitem#1\endcsname}%
\let\auto@bib@innerbib\@empty
%</preamble>
\bibitem [{\citenamefont {Castelvecchi}\ and\ \citenamefont
  {Witze}(2016)}]{Castelvecchi2016}%
  \BibitemOpen
  \bibfield  {author} {\bibinfo {author} {\bibfnamefont {D.}~\bibnamefont
  {Castelvecchi}}\ and\ \bibinfo {author} {\bibfnamefont {A.}~\bibnamefont
  {Witze}},\ }\href@noop {} {\bibfield  {journal} {\bibinfo  {journal} {Nature
  news}\ }\textbf {\bibinfo {volume} {19361}} (\bibinfo {year}
  {2016})}\BibitemShut {NoStop}%
\bibitem [{\citenamefont {Abbott}\ \emph {et~al.}(2016)\citenamefont {Abbott},
  \citenamefont {Abbott}, \citenamefont {Abbott}, \citenamefont {Abernathy},
  \citenamefont {Acernese}, \citenamefont {Ackley}, \citenamefont {Adams},
  \citenamefont {Adams}, \citenamefont {Addesso}, \citenamefont {Adhikari}
  \emph {et~al.}}]{Abbott2016}%
  \BibitemOpen
  \bibfield  {author} {\bibinfo {author} {\bibfnamefont {B.~P.}\ \bibnamefont
  {Abbott}}, \bibinfo {author} {\bibfnamefont {R.}~\bibnamefont {Abbott}},
  \bibinfo {author} {\bibfnamefont {T.~D.}\ \bibnamefont {Abbott}}, \bibinfo
  {author} {\bibfnamefont {M.~R.}\ \bibnamefont {Abernathy}}, \bibinfo {author}
  {\bibfnamefont {F.}~\bibnamefont {Acernese}}, \bibinfo {author}
  {\bibfnamefont {K.}~\bibnamefont {Ackley}}, \bibinfo {author} {\bibfnamefont
  {C.}~\bibnamefont {Adams}}, \bibinfo {author} {\bibfnamefont
  {T.}~\bibnamefont {Adams}}, \bibinfo {author} {\bibfnamefont
  {P.}~\bibnamefont {Addesso}}, \bibinfo {author} {\bibfnamefont {R.~X.}\
  \bibnamefont {Adhikari}}, \emph {et~al.},\ }\href@noop {} {\bibfield
  {journal} {\bibinfo  {journal} {Physical review letters}\ }\textbf {\bibinfo
  {volume} {116}},\ \bibinfo {pages} {061102} (\bibinfo {year}
  {2016})}\BibitemShut {NoStop}%
\bibitem [{\citenamefont {Danzmann}\ \emph {et~al.}(1996)\citenamefont
  {Danzmann}, \citenamefont {Team} \emph {et~al.}}]{Danzmann1996}%
  \BibitemOpen
  \bibfield  {author} {\bibinfo {author} {\bibfnamefont {K.}~\bibnamefont
  {Danzmann}}, \bibinfo {author} {\bibfnamefont {L.~S.}\ \bibnamefont {Team}},
  \emph {et~al.},\ }\href@noop {} {\bibfield  {journal} {\bibinfo  {journal}
  {Classical and Quantum Gravity}\ }\textbf {\bibinfo {volume} {13}},\ \bibinfo
  {pages} {A247} (\bibinfo {year} {1996})}\BibitemShut {NoStop}%
\bibitem [{\citenamefont {Hu}\ and\ \citenamefont {Wu}(2017)}]{Hu2017}%
  \BibitemOpen
  \bibfield  {author} {\bibinfo {author} {\bibfnamefont {W.-R.}\ \bibnamefont
  {Hu}}\ and\ \bibinfo {author} {\bibfnamefont {Y.-L.}\ \bibnamefont {Wu}},\
  }\href@noop {} {\bibinfo {title} {The taiji program in space for
  gravitational wave physics and the nature of gravity}} (\bibinfo {year}
  {2017})\BibitemShut {NoStop}%
\bibitem [{\citenamefont {Eardley}\ \emph {et~al.}(1973)\citenamefont
  {Eardley}, \citenamefont {Lee},\ and\ \citenamefont
  {Lightman}}]{Eardley1973}%
  \BibitemOpen
  \bibfield  {author} {\bibinfo {author} {\bibfnamefont {D.~M.}\ \bibnamefont
  {Eardley}}, \bibinfo {author} {\bibfnamefont {D.~L.}\ \bibnamefont {Lee}},\
  and\ \bibinfo {author} {\bibfnamefont {A.~P.}\ \bibnamefont {Lightman}},\
  }\href@noop {} {\bibfield  {journal} {\bibinfo  {journal} {Physical Review
  D}\ }\textbf {\bibinfo {volume} {8}},\ \bibinfo {pages} {3308} (\bibinfo
  {year} {1973})}\BibitemShut {NoStop}%
\bibitem [{\citenamefont {Wu}(2016)}]{Wu2016}%
  \BibitemOpen
  \bibfield  {author} {\bibinfo {author} {\bibfnamefont {Y.-L.}\ \bibnamefont
  {Wu}},\ }\href {https://doi.org/10.1103/PhysRevD.93.024012} {\bibfield
  {journal} {\bibinfo  {journal} {Phys. Rev. D}\ }\textbf {\bibinfo {volume}
  {93}},\ \bibinfo {pages} {024012} (\bibinfo {year} {2016})},\ \Eprint
  {https://arxiv.org/abs/1506.01807} {arXiv:1506.01807 [hep-th]} \BibitemShut
  {NoStop}%
\bibitem [{\citenamefont {Wu}(2018)}]{Wu:2017urh}%
  \BibitemOpen
  \bibfield  {author} {\bibinfo {author} {\bibfnamefont {Y.-L.}\ \bibnamefont
  {Wu}},\ }\href {https://doi.org/10.1140/epjc/s10052-017-5504-3} {\bibfield
  {journal} {\bibinfo  {journal} {Eur. Phys. J. C}\ }\textbf {\bibinfo {volume}
  {78}},\ \bibinfo {pages} {28} (\bibinfo {year} {2018})},\ \Eprint
  {https://arxiv.org/abs/1712.04537} {arXiv:1712.04537 [hep-th]} \BibitemShut
  {NoStop}%
\bibitem [{\citenamefont {Wu}(2023)}]{Wu2023}%
  \BibitemOpen
  \bibfield  {author} {\bibinfo {author} {\bibfnamefont {Y.-L.}\ \bibnamefont
  {Wu}},\ }\href@noop {} {\bibfield  {journal} {\bibinfo  {journal} {Science
  China Physics, Mechanics \& Astronomy}\ }\textbf {\bibinfo {volume} {66}},\
  \bibinfo {pages} {260411} (\bibinfo {year} {2023})}\BibitemShut {NoStop}%
\bibitem [{\citenamefont {Wu}(2025{\natexlab{a}})}]{Wu:2024mul}%
  \BibitemOpen
  \bibfield  {author} {\bibinfo {author} {\bibfnamefont {Y.-L.}\ \bibnamefont
  {Wu}},\ }\href {https://doi.org/10.1016/j.physletb.2025.139689} {\bibfield
  {journal} {\bibinfo  {journal} {Phys. Lett. B}\ }\textbf {\bibinfo {volume}
  {868}},\ \bibinfo {pages} {139689} (\bibinfo {year} {2025}{\natexlab{a}})},\
  \Eprint {https://arxiv.org/abs/2411.15166} {arXiv:2411.15166
  [physics.gen-ph]} \BibitemShut {NoStop}%
\bibitem [{\citenamefont {Wu}(2025{\natexlab{b}})}]{Wu:2025abi}%
  \BibitemOpen
  \bibfield  {author} {\bibinfo {author} {\bibfnamefont {Y.-L.}\ \bibnamefont
  {Wu}},\ }\href {https://doi.org/10.1016/j.scib.2025.03.026} {\bibfield
  {journal} {\bibinfo  {journal} {Sci. Bull.}\ }\textbf {\bibinfo {volume}
  {70}},\ \bibinfo {pages} {1740} (\bibinfo {year} {2025}{\natexlab{b}})},\
  \Eprint {https://arxiv.org/abs/2502.19458} {arXiv:2502.19458
  [physics.gen-ph]} \BibitemShut {NoStop}%
\bibitem [{\citenamefont {Wu}(2026)}]{Wu:2025rei}%
  \BibitemOpen
  \bibfield  {author} {\bibinfo {author} {\bibfnamefont {Y.-L.}\ \bibnamefont
  {Wu}},\ }\href {https://doi.org/10.1007/s11433-025-2876-0} {\bibfield
  {journal} {\bibinfo  {journal} {Sci. China Phys. Mech. Astron.}\ }\textbf
  {\bibinfo {volume} {69}},\ \bibinfo {pages} {241011} (\bibinfo {year}
  {2026})},\ \Eprint {https://arxiv.org/abs/2508.20128} {arXiv:2508.20128
  [physics.gen-ph]} \BibitemShut {NoStop}%
\bibitem [{\citenamefont {Gao}\ \emph {et~al.}(2024)\citenamefont {Gao},
  \citenamefont {Huang}, \citenamefont {Ma}, \citenamefont {Tang},
  \citenamefont {Wu},\ and\ \citenamefont {Zhou}}]{Gao2024}%
  \BibitemOpen
  \bibfield  {author} {\bibinfo {author} {\bibfnamefont {Y.-K.}\ \bibnamefont
  {Gao}}, \bibinfo {author} {\bibfnamefont {D.}~\bibnamefont {Huang}}, \bibinfo
  {author} {\bibfnamefont {Y.-L.}\ \bibnamefont {Ma}}, \bibinfo {author}
  {\bibfnamefont {Y.}~\bibnamefont {Tang}}, \bibinfo {author} {\bibfnamefont
  {Y.-L.}\ \bibnamefont {Wu}},\ and\ \bibinfo {author} {\bibfnamefont {Y.-F.}\
  \bibnamefont {Zhou}},\ }\href {https://doi.org/10.1103/PhysRevD.109.064072}
  {\bibfield  {journal} {\bibinfo  {journal} {Phys. Rev. D}\ }\textbf {\bibinfo
  {volume} {109}},\ \bibinfo {pages} {064072} (\bibinfo {year} {2024})},\
  \Eprint {https://arxiv.org/abs/2403.17619} {arXiv:2403.17619 [gr-qc]}
  \BibitemShut {NoStop}%
\bibitem [{\citenamefont {Gao}\ \emph {et~al.}(2025)\citenamefont {Gao},
  \citenamefont {Huang},\ and\ \citenamefont {Wu}}]{Gao2025}%
  \BibitemOpen
  \bibfield  {author} {\bibinfo {author} {\bibfnamefont {Y.-K.}\ \bibnamefont
  {Gao}}, \bibinfo {author} {\bibfnamefont {D.}~\bibnamefont {Huang}},\ and\
  \bibinfo {author} {\bibfnamefont {Y.-L.}\ \bibnamefont {Wu}},\ }\href@noop {}
  {\bibfield  {journal} {\bibinfo  {journal} {The European Physical Journal C}\
  }\textbf {\bibinfo {volume} {85}},\ \bibinfo {pages} {1159} (\bibinfo {year}
  {2025})}\BibitemShut {NoStop}%
\bibitem [{\citenamefont {Xu}\ \emph {et~al.}(2025{\natexlab{a}})\citenamefont
  {Xu}, \citenamefont {Jin},\ and\ \citenamefont {Wu}}]{Xu2025a}%
  \BibitemOpen
  \bibfield  {author} {\bibinfo {author} {\bibfnamefont {C.}~\bibnamefont
  {Xu}}, \bibinfo {author} {\bibfnamefont {H.-B.}\ \bibnamefont {Jin}},\ and\
  \bibinfo {author} {\bibfnamefont {Y.-L.}\ \bibnamefont {Wu}},\ }\href@noop {}
  {\bibfield  {journal} {\bibinfo  {journal} {arXiv preprint arXiv:2504.01809}\
  } (\bibinfo {year} {2025}{\natexlab{a}})}\BibitemShut {NoStop}%
\bibitem [{\citenamefont {Amaro-Seoane}\ \emph {et~al.}(2017)\citenamefont
  {Amaro-Seoane}, \citenamefont {Audley}, \citenamefont {Babak}, \citenamefont
  {Baker}, \citenamefont {Barausse}, \citenamefont {Bender}, \citenamefont
  {Berti}, \citenamefont {Binetruy}, \citenamefont {Born}, \citenamefont
  {Bortoluzzi} \emph {et~al.}}]{AmaroSeoane2017}%
  \BibitemOpen
  \bibfield  {author} {\bibinfo {author} {\bibfnamefont {P.}~\bibnamefont
  {Amaro-Seoane}}, \bibinfo {author} {\bibfnamefont {H.}~\bibnamefont
  {Audley}}, \bibinfo {author} {\bibfnamefont {S.}~\bibnamefont {Babak}},
  \bibinfo {author} {\bibfnamefont {J.}~\bibnamefont {Baker}}, \bibinfo
  {author} {\bibfnamefont {E.}~\bibnamefont {Barausse}}, \bibinfo {author}
  {\bibfnamefont {P.}~\bibnamefont {Bender}}, \bibinfo {author} {\bibfnamefont
  {E.}~\bibnamefont {Berti}}, \bibinfo {author} {\bibfnamefont
  {P.}~\bibnamefont {Binetruy}}, \bibinfo {author} {\bibfnamefont
  {M.}~\bibnamefont {Born}}, \bibinfo {author} {\bibfnamefont {D.}~\bibnamefont
  {Bortoluzzi}}, \emph {et~al.},\ }\href@noop {} {\bibfield  {journal}
  {\bibinfo  {journal} {arXiv preprint arXiv:1702.00786}\ } (\bibinfo {year}
  {2017})}\BibitemShut {NoStop}%
\bibitem [{\citenamefont {Tinto}\ and\ \citenamefont
  {Larson}(2004)}]{Tinto:2004nz}%
  \BibitemOpen
  \bibfield  {author} {\bibinfo {author} {\bibfnamefont {M.}~\bibnamefont
  {Tinto}}\ and\ \bibinfo {author} {\bibfnamefont {S.~L.}\ \bibnamefont
  {Larson}},\ }\href {https://doi.org/10.1103/PhysRevD.70.062002} {\bibfield
  {journal} {\bibinfo  {journal} {Phys. Rev. D}\ }\textbf {\bibinfo {volume}
  {70}},\ \bibinfo {pages} {062002} (\bibinfo {year} {2004})},\ \Eprint
  {https://arxiv.org/abs/gr-qc/0405147} {arXiv:gr-qc/0405147} \BibitemShut
  {NoStop}%
\bibitem [{\citenamefont {Barroso}\ \emph {et~al.}(2025)\citenamefont
  {Barroso}, \citenamefont {Lemi{\`e}re}, \citenamefont {Mauger},\ and\
  \citenamefont {Baghi}}]{Barroso:2024azl}%
  \BibitemOpen
  \bibfield  {author} {\bibinfo {author} {\bibfnamefont {R.~C.}\ \bibnamefont
  {Barroso}}, \bibinfo {author} {\bibfnamefont {Y.}~\bibnamefont
  {Lemi{\`e}re}}, \bibinfo {author} {\bibfnamefont {F.}~\bibnamefont
  {Mauger}},\ and\ \bibinfo {author} {\bibfnamefont {Q.}~\bibnamefont
  {Baghi}},\ }\href {https://doi.org/10.1088/1361-6382/addbbd} {\bibfield
  {journal} {\bibinfo  {journal} {Class. Quant. Grav.}\ }\textbf {\bibinfo
  {volume} {42}},\ \bibinfo {pages} {115016} (\bibinfo {year} {2025})},\
  \Eprint {https://arxiv.org/abs/2406.00190} {arXiv:2406.00190 [gr-qc]}
  \BibitemShut {NoStop}%
\bibitem [{\citenamefont {Xu}\ \emph {et~al.}(2025{\natexlab{b}})\citenamefont
  {Xu}, \citenamefont {Yao}, \citenamefont {Tang},\ and\ \citenamefont
  {Wu}}]{Xu2025}%
  \BibitemOpen
  \bibfield  {author} {\bibinfo {author} {\bibfnamefont {H.-T.}\ \bibnamefont
  {Xu}}, \bibinfo {author} {\bibfnamefont {Y.-H.}\ \bibnamefont {Yao}},
  \bibinfo {author} {\bibfnamefont {Y.}~\bibnamefont {Tang}},\ and\ \bibinfo
  {author} {\bibfnamefont {Y.-L.}\ \bibnamefont {Wu}},\ }\href
  {https://doi.org/10.1103/xlqy-r6n3} {\bibfield  {journal} {\bibinfo
  {journal} {Phys. Rev. D}\ }\textbf {\bibinfo {volume} {112}},\ \bibinfo
  {pages} {095021} (\bibinfo {year} {2025}{\natexlab{b}})},\ \Eprint
  {https://arxiv.org/abs/2506.09744} {arXiv:2506.09744 [hep-ph]} \BibitemShut
  {NoStop}%
\bibitem [{\citenamefont {Cornish}\ and\ \citenamefont
  {Rubbo}(2003)}]{Cornish2003}%
  \BibitemOpen
  \bibfield  {author} {\bibinfo {author} {\bibfnamefont {N.~J.}\ \bibnamefont
  {Cornish}}\ and\ \bibinfo {author} {\bibfnamefont {L.~J.}\ \bibnamefont
  {Rubbo}},\ }\href@noop {} {\bibfield  {journal} {\bibinfo  {journal}
  {Physical Review D}\ }\textbf {\bibinfo {volume} {67}},\ \bibinfo {pages}
  {022001} (\bibinfo {year} {2003})}\BibitemShut {NoStop}%
\bibitem [{\citenamefont {Cornish}\ and\ \citenamefont
  {Hellings}(2003)}]{Cornish2003a}%
  \BibitemOpen
  \bibfield  {author} {\bibinfo {author} {\bibfnamefont {N.~J.}\ \bibnamefont
  {Cornish}}\ and\ \bibinfo {author} {\bibfnamefont {R.~W.}\ \bibnamefont
  {Hellings}},\ }\href@noop {} {\bibfield  {journal} {\bibinfo  {journal}
  {Classical and Quantum Gravity}\ }\textbf {\bibinfo {volume} {20}},\ \bibinfo
  {pages} {4851} (\bibinfo {year} {2003})}\BibitemShut {NoStop}%
\bibitem [{\citenamefont {Tinto}\ and\ \citenamefont
  {Dhurandhar}(2021)}]{Tinto2021}%
  \BibitemOpen
  \bibfield  {author} {\bibinfo {author} {\bibfnamefont {M.}~\bibnamefont
  {Tinto}}\ and\ \bibinfo {author} {\bibfnamefont {S.~V.}\ \bibnamefont
  {Dhurandhar}},\ }\href@noop {} {\bibfield  {journal} {\bibinfo  {journal}
  {Living Reviews in Relativity}\ }\textbf {\bibinfo {volume} {24}},\ \bibinfo
  {pages} {1} (\bibinfo {year} {2021})}\BibitemShut {NoStop}%
\bibitem [{\citenamefont {Armstrong}\ \emph {et~al.}(1999)\citenamefont
  {Armstrong}, \citenamefont {Estabrook},\ and\ \citenamefont
  {Tinto}}]{Armstrong1999}%
  \BibitemOpen
  \bibfield  {author} {\bibinfo {author} {\bibfnamefont {J.}~\bibnamefont
  {Armstrong}}, \bibinfo {author} {\bibfnamefont {F.}~\bibnamefont
  {Estabrook}},\ and\ \bibinfo {author} {\bibfnamefont {M.}~\bibnamefont
  {Tinto}},\ }\href@noop {} {\bibfield  {journal} {\bibinfo  {journal} {The
  Astrophysical Journal}\ }\textbf {\bibinfo {volume} {527}},\ \bibinfo {pages}
  {814} (\bibinfo {year} {1999})}\BibitemShut {NoStop}%
\bibitem [{\citenamefont {Dhurandhar}\ \emph {et~al.}(2002)\citenamefont
  {Dhurandhar}, \citenamefont {Nayak},\ and\ \citenamefont
  {Vinet}}]{Dhurandhar2002}%
  \BibitemOpen
  \bibfield  {author} {\bibinfo {author} {\bibfnamefont {S.}~\bibnamefont
  {Dhurandhar}}, \bibinfo {author} {\bibfnamefont {K.~R.}\ \bibnamefont
  {Nayak}},\ and\ \bibinfo {author} {\bibfnamefont {J.-Y.}\ \bibnamefont
  {Vinet}},\ }\href@noop {} {\bibfield  {journal} {\bibinfo  {journal}
  {Physical Review D}\ }\textbf {\bibinfo {volume} {65}},\ \bibinfo {pages}
  {102002} (\bibinfo {year} {2002})}\BibitemShut {NoStop}%
\bibitem [{\citenamefont {Estabrook}\ \emph {et~al.}(2000)\citenamefont
  {Estabrook}, \citenamefont {Tinto},\ and\ \citenamefont
  {Armstrong}}]{Estabrook2000}%
  \BibitemOpen
  \bibfield  {author} {\bibinfo {author} {\bibfnamefont {F.}~\bibnamefont
  {Estabrook}}, \bibinfo {author} {\bibfnamefont {M.}~\bibnamefont {Tinto}},\
  and\ \bibinfo {author} {\bibfnamefont {J.}~\bibnamefont {Armstrong}},\
  }\href@noop {} {\bibfield  {journal} {\bibinfo  {journal} {Physical Review
  D}\ }\textbf {\bibinfo {volume} {62}},\ \bibinfo {pages} {042002} (\bibinfo
  {year} {2000})}\BibitemShut {NoStop}%
\bibitem [{\citenamefont {Vallisneri}(2005)}]{Vallisneri2005}%
  \BibitemOpen
  \bibfield  {author} {\bibinfo {author} {\bibfnamefont {M.}~\bibnamefont
  {Vallisneri}},\ }\href@noop {} {\bibfield  {journal} {\bibinfo  {journal}
  {Physical Review D—Particles, Fields, Gravitation, and Cosmology}\ }\textbf
  {\bibinfo {volume} {72}},\ \bibinfo {pages} {042003} (\bibinfo {year}
  {2005})}\BibitemShut {NoStop}%
\bibitem [{\citenamefont {Tinto}\ \emph {et~al.}(2023)\citenamefont {Tinto},
  \citenamefont {Dhurandhar},\ and\ \citenamefont {Malakar}}]{Tinto:2022zmf}%
  \BibitemOpen
  \bibfield  {author} {\bibinfo {author} {\bibfnamefont {M.}~\bibnamefont
  {Tinto}}, \bibinfo {author} {\bibfnamefont {S.}~\bibnamefont {Dhurandhar}},\
  and\ \bibinfo {author} {\bibfnamefont {D.}~\bibnamefont {Malakar}},\ }\href
  {https://doi.org/10.1103/PhysRevD.107.082001} {\bibfield  {journal} {\bibinfo
   {journal} {Phys. Rev. D}\ }\textbf {\bibinfo {volume} {107}},\ \bibinfo
  {pages} {082001} (\bibinfo {year} {2023})},\ \Eprint
  {https://arxiv.org/abs/2212.05967} {arXiv:2212.05967 [gr-qc]} \BibitemShut
  {NoStop}%
\bibitem [{\citenamefont {Yu}\ \emph {et~al.}(2023)\citenamefont {Yu},
  \citenamefont {Yao}, \citenamefont {Tang},\ and\ \citenamefont
  {Wu}}]{Yu:2023iog}%
  \BibitemOpen
  \bibfield  {author} {\bibinfo {author} {\bibfnamefont {J.-C.}\ \bibnamefont
  {Yu}}, \bibinfo {author} {\bibfnamefont {Y.-H.}\ \bibnamefont {Yao}},
  \bibinfo {author} {\bibfnamefont {Y.}~\bibnamefont {Tang}},\ and\ \bibinfo
  {author} {\bibfnamefont {Y.-L.}\ \bibnamefont {Wu}},\ }\href
  {https://doi.org/10.1103/PhysRevD.108.083007} {\bibfield  {journal} {\bibinfo
   {journal} {Phys. Rev. D}\ }\textbf {\bibinfo {volume} {108}},\ \bibinfo
  {pages} {083007} (\bibinfo {year} {2023})},\ \Eprint
  {https://arxiv.org/abs/2307.09197} {arXiv:2307.09197 [gr-qc]} \BibitemShut
  {NoStop}%
\bibitem [{\citenamefont {Babak}\ \emph {et~al.}(2021)\citenamefont {Babak},
  \citenamefont {Hewitson},\ and\ \citenamefont {Petiteau}}]{Stanislav2021}%
  \BibitemOpen
  \bibfield  {author} {\bibinfo {author} {\bibfnamefont {S.}~\bibnamefont
  {Babak}}, \bibinfo {author} {\bibfnamefont {M.}~\bibnamefont {Hewitson}},\
  and\ \bibinfo {author} {\bibfnamefont {A.}~\bibnamefont {Petiteau}},\
  }\href@noop {} {\bibinfo {title} {Lisa sensitivity and snr calculations}}
  (\bibinfo {year} {2021})\BibitemShut {NoStop}%
\bibitem [{\citenamefont {Rubbo}\ \emph {et~al.}(2004)\citenamefont {Rubbo},
  \citenamefont {Cornish},\ and\ \citenamefont {Poujade}}]{Rubbo2004}%
  \BibitemOpen
  \bibfield  {author} {\bibinfo {author} {\bibfnamefont {L.~J.}\ \bibnamefont
  {Rubbo}}, \bibinfo {author} {\bibfnamefont {N.~J.}\ \bibnamefont {Cornish}},\
  and\ \bibinfo {author} {\bibfnamefont {O.}~\bibnamefont {Poujade}},\
  }\href@noop {} {\bibfield  {journal} {\bibinfo  {journal} {Physical Review
  D}\ }\textbf {\bibinfo {volume} {69}},\ \bibinfo {pages} {082003} (\bibinfo
  {year} {2004})}\BibitemShut {NoStop}%
\bibitem [{\citenamefont {Shi}\ and\ \citenamefont
  {Tang}(2026)}]{shi2026construction}%
  \BibitemOpen
  \bibfield  {author} {\bibinfo {author} {\bibfnamefont {H.-Y.}\ \bibnamefont
  {Shi}}\ and\ \bibinfo {author} {\bibfnamefont {Y.}~\bibnamefont {Tang}},\
  }\href@noop {} {\bibfield  {journal} {\bibinfo  {journal} {arXiv preprint
  arXiv:2606.28071}\ } (\bibinfo {year} {2026})}\BibitemShut {NoStop}%
\bibitem [{\citenamefont {Lin}\ \emph {et~al.}(2024)\citenamefont {Lin},
  \citenamefont {Yu},\ and\ \citenamefont {Gong}}]{lin2024limiting}%
  \BibitemOpen
  \bibfield  {author} {\bibinfo {author} {\bibfnamefont {Z.-C.}\ \bibnamefont
  {Lin}}, \bibinfo {author} {\bibfnamefont {H.}~\bibnamefont {Yu}},\ and\
  \bibinfo {author} {\bibfnamefont {Y.}~\bibnamefont {Gong}},\ }\href@noop {}
  {\bibfield  {journal} {\bibinfo  {journal} {Physical Review D}\ }\textbf
  {\bibinfo {volume} {109}},\ \bibinfo {pages} {104015} (\bibinfo {year}
  {2024})}\BibitemShut {NoStop}%
\bibitem [{\citenamefont {Liang}\ \emph {et~al.}(2026)\citenamefont {Liang},
  \citenamefont {Huang}, \citenamefont {Zhang},\ and\ \citenamefont
  {Hu}}]{liang2026probing}%
  \BibitemOpen
  \bibfield  {author} {\bibinfo {author} {\bibfnamefont {Z.-C.}\ \bibnamefont
  {Liang}}, \bibinfo {author} {\bibfnamefont {F.-P.}\ \bibnamefont {Huang}},
  \bibinfo {author} {\bibfnamefont {X.}~\bibnamefont {Zhang}},\ and\ \bibinfo
  {author} {\bibfnamefont {Y.-M.}\ \bibnamefont {Hu}},\ }\href@noop {}
  {\bibfield  {journal} {\bibinfo  {journal} {Chinese Physics C}\ }\textbf
  {\bibinfo {volume} {50}},\ \bibinfo {pages} {035104} (\bibinfo {year}
  {2026})}\BibitemShut {NoStop}%
\bibitem [{\citenamefont {Wu}\ \emph {et~al.}(2024)\citenamefont {Wu},
  \citenamefont {Li}, \citenamefont {Liu},\ and\ \citenamefont
  {Cao}}]{wu2024comparison}%
  \BibitemOpen
  \bibfield  {author} {\bibinfo {author} {\bibfnamefont {J.}~\bibnamefont
  {Wu}}, \bibinfo {author} {\bibfnamefont {J.}~\bibnamefont {Li}}, \bibinfo
  {author} {\bibfnamefont {X.}~\bibnamefont {Liu}},\ and\ \bibinfo {author}
  {\bibfnamefont {Z.}~\bibnamefont {Cao}},\ }\href@noop {} {\bibfield
  {journal} {\bibinfo  {journal} {Physical Review D}\ }\textbf {\bibinfo
  {volume} {109}},\ \bibinfo {pages} {104014} (\bibinfo {year}
  {2024})}\BibitemShut {NoStop}%
\bibitem [{\citenamefont {Gao}\ \emph {et~al.}(2023)\citenamefont {Gao},
  \citenamefont {You}, \citenamefont {Gong}, \citenamefont {Zhang},\ and\
  \citenamefont {Zhang}}]{gao2023testing}%
  \BibitemOpen
  \bibfield  {author} {\bibinfo {author} {\bibfnamefont {Q.}~\bibnamefont
  {Gao}}, \bibinfo {author} {\bibfnamefont {Y.}~\bibnamefont {You}}, \bibinfo
  {author} {\bibfnamefont {Y.}~\bibnamefont {Gong}}, \bibinfo {author}
  {\bibfnamefont {C.}~\bibnamefont {Zhang}},\ and\ \bibinfo {author}
  {\bibfnamefont {C.}~\bibnamefont {Zhang}},\ }\href@noop {} {\bibfield
  {journal} {\bibinfo  {journal} {Physical Review D}\ }\textbf {\bibinfo
  {volume} {108}},\ \bibinfo {pages} {024027} (\bibinfo {year}
  {2023})}\BibitemShut {NoStop}%
\bibitem [{\citenamefont {Zhou}\ \emph {et~al.}(2025)\citenamefont {Zhou},
  \citenamefont {Wang},\ and\ \citenamefont {Shao}}]{zhou2025sensitivity}%
  \BibitemOpen
  \bibfield  {author} {\bibinfo {author} {\bibfnamefont {J.}~\bibnamefont
  {Zhou}}, \bibinfo {author} {\bibfnamefont {P.-P.}\ \bibnamefont {Wang}},\
  and\ \bibinfo {author} {\bibfnamefont {C.-G.}\ \bibnamefont {Shao}},\
  }\href@noop {} {\bibfield  {journal} {\bibinfo  {journal} {Classical and
  Quantum Gravity}\ }\textbf {\bibinfo {volume} {42}},\ \bibinfo {pages}
  {025013} (\bibinfo {year} {2025})}\BibitemShut {NoStop}%
\bibitem [{\citenamefont {Lu}\ \emph {et~al.}(2026)\citenamefont {Lu},
  \citenamefont {Lin}, \citenamefont {Zhu}, \citenamefont {Liu},\ and\
  \citenamefont {Zhang}}]{lu2026gravitational}%
  \BibitemOpen
  \bibfield  {author} {\bibinfo {author} {\bibfnamefont {S.}~\bibnamefont
  {Lu}}, \bibinfo {author} {\bibfnamefont {H.-J.}\ \bibnamefont {Lin}},
  \bibinfo {author} {\bibfnamefont {T.}~\bibnamefont {Zhu}}, \bibinfo {author}
  {\bibfnamefont {Y.-X.}\ \bibnamefont {Liu}},\ and\ \bibinfo {author}
  {\bibfnamefont {X.}~\bibnamefont {Zhang}},\ }\href@noop {} {\bibfield
  {journal} {\bibinfo  {journal} {The European Physical Journal C}\ }\textbf
  {\bibinfo {volume} {86}},\ \bibinfo {pages} {283} (\bibinfo {year}
  {2026})}\BibitemShut {NoStop}%
\bibitem [{\citenamefont {Hartwig}\ \emph {et~al.}(2023)\citenamefont
  {Hartwig}, \citenamefont {Lilley}, \citenamefont {Muratore},\ and\
  \citenamefont {Pieroni}}]{Hartwig2023}%
  \BibitemOpen
  \bibfield  {author} {\bibinfo {author} {\bibfnamefont {O.}~\bibnamefont
  {Hartwig}}, \bibinfo {author} {\bibfnamefont {M.}~\bibnamefont {Lilley}},
  \bibinfo {author} {\bibfnamefont {M.}~\bibnamefont {Muratore}},\ and\
  \bibinfo {author} {\bibfnamefont {M.}~\bibnamefont {Pieroni}},\ }\href@noop
  {} {\bibfield  {journal} {\bibinfo  {journal} {Physical Review D}\ }\textbf
  {\bibinfo {volume} {107}},\ \bibinfo {pages} {123531} (\bibinfo {year}
  {2023})}\BibitemShut {NoStop}%
\end{thebibliography}%
